\documentstyle[11pt,aaspp4]{article}

\lefthead{Calzetti}
\righthead{HST-WFPC2 Imaging of NGC5253}

\begin{document}

\title{Dust and Recent Star Formation\\ in the Core of NGC5253.
\altaffilmark{1}}

\author{Daniela Calzetti}
\affil{Space Telescope Science Institute, 3700 San Martin Dr., 
   Baltimore, MD 21218, USA; e-mail: calzetti@stsci.edu}
\author{Gerhardt R. Meurer}
\affil{Dept. of Physics and Astronomy, The Johns Hopkins University}
\author{Ralph C. Bohlin}
\affil{Space Telescope Science Institute}
\author{Donald R. Garnett}
\affil{Dept. of Astronomy, University of Minnesota}
\author{Anne L. Kinney and Claus Leitherer}
\affil{Space Telescope Science Institute}
\and
\author{Thaisa Storchi-Bergmann}
\affil{Instituto de Fisica, Universidade Federal Rio Grande do Sul}

\altaffiltext{1}{Based on observations obtained with the NASA/ESA {\it Hubble 
Space Telescope} at the Space Telescope Science Institute, which is operated 
by the Association of Universities for Research in Astronomy, Inc., under 
NASA contract NAS5-26555.} 

\begin{abstract}

Ultraviolet and optical narrow and broad band images of NGC5253
obtained with the Hubble Space Telescope Wide Field and Planetary
Camera 2 are used to derive the properties of the dust distribution
and the recent star formation history of this metal-poor dwarf galaxy.
Corrections for the effects of dust are important in the center of
NGC5253: dust reddening is markedly inhomogeneous across the galaxy's
central 20\arcsec~ region. One of the most obscured regions coincides
with the region of highest star formation activity in the galaxy;
clouds of more than 9~magnitudes of optical depth at V enshroud a
2.5~Myr old stellar cluster in this area. The ages of the bright
clusters in the center of the galaxy are anti-correlated with the
amount of dust obscuration the cluster suffers. This result agrees
with the expectation that young stellar associations are located in
heavily obscured regions, but after only 2--3~Myr they remove/emerge
from the parental dust cloud and become almost extinction-free. On
average, the continuum emission of the diffuse stellar population is
about a factor two less reddened than the ionized gas emission, a
behavior typical of starburst galaxies (Calzetti et al. 1994). In the
case of NGC5253, this difference originates from the larger scale
length of the star distribution relative to the ionized gas: the half
light radius of the UV-bright stars is about twice as large as the
half light radius of the ionized gas emission.

Star formation has been active at least over the past 100~Myr in the
central 20\arcsec~ of the galaxy, as indicated by the age distribution
of both the blue diffuse stellar population and the bright stellar
clusters. The star formation episodes may have been discrete in time, 
or almost continuous but variable in intensity and spatial
extension. The current peak of the star formation is located in a
6\arcsec~ region, more spatially concentrated than the star
formation averaged over the past 100~Myr. Its average star formation
intensity is 10$^{-5}$--10$^{-4}$~M$_{\odot}$/yr/pc$^2$ for a
0.1--100~M$_{\odot}$ Salpeter IMF, a factor 10 to 100 times larger
than in the galaxy's central 20\arcsec. This starburst region contains
a stellar population $\sim$5~Myr old and the two youngest (2.5~Myr and
$\sim$3--4~Myr, respectively) of the bright stellar clusters in the
galaxy's center. The two clusters contribute between 20\% and 65\% of
the ionizing photons in the starburst, a contribution between 1.3 and
4.3 times larger than the average over the central 20~\arcsec. This is
expected if cluster formation is an important mode of star formation
in the early phase of a starburst event. The mass of the 2.5~Myr old
cluster may be as large as $\sim$10$^6$~M$_{\odot}$, making this one a
Super-Star-Cluster candidate.
\end{abstract}


\section{Introduction}

Bursts of star formation play an important role in the framework of
galaxy evolution. In the local Universe, more than half of the
high-mass star formation comes from the nuclear region of galaxies
(Gallego et al. 1995); within 10~Mpc of the Galaxy, 25\% of the
high-mass star formation is produced by less than a handful of
starburst galaxies (Heckman 1997). At intermediate redshifts
(z$\sim$0.3), the bulk of the excess faint blue galaxy counts may be
due to a dwarf galaxy population experiencing sudden, high-efficiency
and fast-evolving starbursts (e.g., Colless et al. 1994, Phillips \&
Driver 1995, Babul \& Ferguson 1996). At higher redshifts (z$\sim$3),
the recently discovered galaxy population (Steidel et al. 1996) has
properties indicating a high-intensity starforming phase (e.g.,
Steidel et al. 1996, Giavalisco, Steidel \& Macchetto 1996; see, also,
Lowenthal et al. 1997).  To the extent that the initial mass function
(IMF) is independent of the environment (Massey et al. 1995; Moffat
1996; Stas\'inska \& Leitherer 1996), a significant fraction of star
formation in the universe appears to occur through high intensity
episodes.  Hence to understand galaxy evolution, and cosmic evolution,
we must understand bursts of star formation.
 
Open questions about starbursts most relevant to cosmic evolution
include the following.  (1) How long do starbursts last?  The answer,
determines how important individual burst are in shaping the galaxy's
stellar population; is a galaxy likely to undergo short-lived
(1--10~Myr) bursts or long-lived ($\gg$10~Myr) bursts? Bursts duration
also has bearing on the problem of the relative number of burst and
postburst systems (Norman 1991).  (2) How does star formation spread
within a galaxy?  At issue is how starbursts are fueled (e.g. Shlosman
1992, Olson \& Kwan 1990), and whether the energetic feedback of high
mass star formation into the ISM inhibits or enhances star formation
(Shull 1993, Chu \& Kennicutt 1994, Heckman, Armus, \& Miley 1990,
Lehnert \& Heckman 1996).  Finally, (3) how much star formation is
obscured or hidden from view?  The impact of dust obscuration on
optical and ultraviolet surveys of galaxies must be assessed in order
to get a census of the actual star formation in the universe.

The best place to look for answers to these questions are the nearest
starburst galaxies. The high spatial resolution and the high
signal-to-noise multiwavelength information which can be obtained for
these systems allow the detailed investigation of the starburst
`phenomenon'.  NGC5253, a dwarf galaxy in the Centaurus Group at the
distance of 4.1~Mpc (Sandage et al. 1994), is an excellent laboratory
for this study. Although the outer regions of the galaxy have
properties similar to a dwarf elliptical galaxy (Sersic, Carranza \&
Pastoriza 1972, Caldwell \& Phillips 1989), star formation is active
in its central 20\arcsec\ (the ``core'' hereafter), as characterized
by a high surface brightness blue continuum and intense line emission
(e.g., Storchi-Bergmann, Kinney \& Challis 1995).  The core hosts
about a dozen blue stellar clusters as well as diffusely distributed
high mass stars (Meurer et al. 1995).  The peak intensity of star
formation, as inferred from the distribution of the H$\alpha$
emission, is concentrated in a $\sim$6\arcsec\ region in the Northern
part of the core (the ``nucleus'' hereafter). The most recent star
formation episode is probably responsible for heating the gas which
emits in the soft X-ray (Martin \& Kennicutt 1995), as well as for the
complex system of loops and filaments of ionized gas (Marlowe et
al. 1995). The inhomogeneous structure and the intense star formation
in the central region of NGC5253 has suggested a classification closer
to an amorphous galaxy or a blue compact dwarf (Sandage \& Brucato
1979, Meurer et al. 1994), rather than an elliptical.

An encounter with the spiral M83, located at a projected separation of
only 130~kpc, about 1--2~Gyr ago has been suggested as the trigger of
the star formation in NGC~5253 (Rogstad, Lockart \& Wright 1974, van
den Bergh 1980, Caldwell \& Phillips 1989). However, the starburst in
the nucleus has probably an age $\lesssim 10$ Myr, as implied by the
presence of Wolf-Rayet stars (Campbell, Terlevich \& Melnick 1986,
Walsh \& Roy 1987, 1989, Schaerer et al. 1997), the small number of
red supergiants (Campbell \& Terlevich 1984), and the purely thermal
component of the radio emission (Beck et al. 1996).  NGC~5253 is
chemically inhomogeneous, displaying a region of enhanced Nitrogen
abundance (Walsh \& Roy 1989, Kobulnicky et al. 1997) which indicate
recent pollution from ejecta of massive stars.  Radial inflow of
peripheral HI gas has been suggested as a possible source of fuel for
the current starburst (Kobulnicky \& Skillman 1995, Turner, Beck \&
Hurt 1997).

NGC~5253 is also an ideal target for mapping the effects of dust on a
starburst.  The extinction is irregular in the core of the galaxy and
the central region is traversed by a dust lane along the E-W
direction, bisecting the galaxy along the minor axis (Walsh \& Roy
1989). The nucleus of NGC5253 is very red above 2~$\mu$m (Moorwood \&
Glass 1982), indicating presence of hot dust probably heated by an
obscured nucleus (A$_V$=8-25~mag, Aitken et al. 1982, Roche et
al. 1991, Telesco, Dressel, \& Wolstencroft 1993).  Despite the
strength of the mid and far infrared emission, the reddening
determined from the intense nuclear Balmer lines of NGC5253 is
generally small, while the ultraviolet (UV) spectral energy
distribution is redder than expected from a young, unextincted
starburst (Gonzalez-Riestra, Rego, \& Zamorano 1987, Kinney et
al. 1993, Calzetti, Kinney \& Storchi-Bergmann 1994, CKS94 hereafter).

The purpose of this paper is twofold. (1) Investigate the past 100~Myr
or so of the star formation history of NGC5253. This timescale can be
considered a sort of ``boundary'' between short-lived and long-lived
bursts of star formation, because it is much longer than the typical
lifetime of O and B stars ($\approx$10~Myr). (2) Measure the impact of
dust obscuration on the intrinsic properties of the galaxy. The
galaxy's star formation history can be tackled via the age
distribution of the stellar populations, which is derived from
multiwavelength observations. Here the effects of reddening must be
carefully measured, since dust reddening and stellar population ageing
produce similar colors, and the spectral energy distribution of a
young dusty stellar population may resemble that of an old dust-free
population.  The first part of the paper is thus devoted at solving
the age/dust degeneracy by constraining reddening effects. The present
dataset is composed of ultraviolet (UV) and optical narrow- and
broad-band images of NGC5253 recently obtained with the Hubble Space
Telescope Wide Field and Planetary Camera~2 (WFPC2). The WFPC2 offers
the high spatial resolution necessary to isolate structures on the
scale of HII complexes in nearby galaxies. The multiwavelength, UV to
I, coverage of the images offers the baseline necessary to determine
accurate ages of the stellar populations from colors.

The paper is organized as follows. The data are presented in Section~2
and the observed morphology of the ionized gas and of the stellar
continuum is presented in Section~3. Section~4 investigates the
effects of dust reddening on the photometric quantities; the results
from this section are used in Section~5 to break the age/reddening
degeneracy and derive the ages of the stellar populations and the
recent star formation history of the galaxy. The summary is given in
Section~6.

\section{The Observations and the Data Reduction}

NGC5253 was observed on May 8-9, 1996 with the HST WFPC2, in the broad
band filters F547M and F814W and in the narrow band filters F487N and
F656N. The log of the exposures, together with the exposure times and
the characteristics of the filters are listed in Table~1. The
starforming core of the galaxy is about 20\arcsec--30\arcsec~ in
diameter and was centered on the WF3 chip
(80$^{\prime\prime}\times$80$^{\prime\prime}$).  A telescope
orientation of about 85 degrees was requested to match the position
and orientation of archival WFPC2 images of the same region in the
F255W filter (GO program \# 6124, P.I. Robert Fesen); a homogeneous
set of data covering the UV and optical wavelength range was therefore
obtained. The final positional displacement of our images relative to
the archival images is only a few pixels, and the relative rotation is
0.25 degrees.

Two pairs of images were obtained in each of the optical filters, with
the second pair shifted by (5,5) pixels relative to the first pair
(cf. Table~1). The F814W filter is the WFPC2 I band; the F547M filter
is used here as a V band equivalent, and was preferred to the wider
F555W filter because its bandpass excludes the strong [OIII](5007~\AA)
nebular emission from the galaxy.  The recession velocity of NGC5253
is sufficiently small (404 km\,s$^{-1}$) that the two narrow band
filters are almost centered at the wavelength of the hydrogen
recombination emission lines H$\beta$(4861~\AA) and
H$\alpha$(6563~\AA), respectively.  The F656N filter is narrow enough
to exclude the redshifted [NII](6584~\AA) emission line, while the
[NII](6548~\AA) emission line represents less than 2-3\% of
the H$\alpha$ flux in this low metallicity galaxy, as inferred from
spectroscopic data (Storchi-Bergmann, Calzetti, \& Kinney 1994).  The
archival dataset includes nine images obtained in the ultraviolet
F255W filter on May 29, 1995. The images are organized in three
triplets, with the second and the third triplet shifted by ($-$5,5)
and (5,5) pixels relative to the first (see Table~1).

The data, both new and archival, were reduced by the STScI calibration
pipeline, which includes flagging of bad pixels, A/D conversion, bias
and dark current subtraction, flatfielding. We repeated the flatfield
procedure on the optical images using the new flatfields available as
of July 1996. The differences in the WF3 between old and new
flatfields is of the order of 2\% or less (peak-to-peak).  The reduced
images were registered to a common position using linear shifts;
cosmic ray and hot pixel rejection and co-addition were performed
using the STSDAS routine CRREJ (Williams et al. 1996), with a
rejection threshold of 4~$\sigma$ for the cosmic rays and 2.4~$\sigma$
for the adjacent pixels. 

The absolute photometric calibration of the images is obtained from
the zero-points listed in HST Data Handbook (1995). The effect of
contaminant buildup onto the WFPC2 window is negligible at optical
wavelengths, but is important in the UV filters. The F255W
observations were obtained 22.8 days after decontamination, with a net
loss in efficiency of about 11\%. The UV images have, therefore, been
corrected for this effect. 

The accuracy of the photometric calibration of the stellar continuum
was checked against the UV (IUE) and optical spectrum of Kinney et
al. (1993) and Storchi-Bergmann et al. (1995). To compare the WFPC2
images with the spectrum, fluxes were extracted from the images using
the same aperture size and orientation of the spectrum. The aperture
was centered at the peak of the optical emission in the images (about
the center of the WF3 chip).  The flux densities as measured from the
images are only a few percent away from the values obtained from the
spectrum convolved with the WFPC2 filters bandpasses (Table~2).  The
photometric zero-point has an accuracy of the order of 2-3\%~ in the
optical filters (HST Data Handbook 1995). The main source of
uncertainty for the optical image-spectrum comparison is the
background subtraction; the galaxy has dimensions 5\farcm00 $\times$
1\farcm95 (calculated at the blue surface brightness level of
25~mag~arcsec$^{-2}$, de Vaucouleurs et al. 1991) and the galaxy's
emission fills both the WF3 chip and the spectroscopic slit.  This
carries an uncertainty of the order of 5--6\% in the background
subtraction for both the V and I bands (see Table~2). A correction of
2\% on the total counts is applied to the UV and the narrow-band
images to correct for the Charge Transfer Efficiency (CTE) problem of
WFPC2 (Holtzmann et al. 1995). The photometric zero-point for the
F255W is known to within 5--7\%, and an additional 3--4\%~
uncertainity comes from the pointing uncertaity of the IUE.

Images of the nebular emission in H$\alpha$ and in H$\beta$ are
obtained by subtracting the stellar continuum from the narrow band
images.  The image of the stellar continuum underlying the H$\alpha$
emission has been obtained by linear interpolation of the F547M and
the F814W images to the central wavelength of the F656N image. The
interpolated image has been multiplied by 1.10 to correct for a small
disagreement in the stellar fluxes between the F656N and the
interpolated images. Because of the galaxy's recession velocity, the
H$\alpha$ emission line is 8.1~\AA~ off-center in the filter bandpass
and a small correction for the filter transmission (+9.4\%) is
applied.

For the stellar continuum underlying the H$\beta$ emission, a linear
extrapolation from the F547M and F814W images is adopted. This
procedure potentially leads to larger uncertainties than for
H$\alpha$, because of differences in the blue continuum in blue and
red stars. An 8\% increase is applied to the extrapolated continuum
images to match the intensity of stars to the level of the F487N
images. A correction for the filter transmission curve (+1.1\%) is
also applied. The effect of the underlying stellar absorption is
larger for the weak H$\beta$ emission than for H$\alpha$. Using red
stellar objects to scale the continuum image to the narrow-band image
can introduce a bias in the continuum subtraction of the blue stars,
depending on the red stellar population. For example, an F/G type
population has a relatively high H$\beta$ absorption equivalent width,
which would lead to undersubtraction of the OB stars continuum; on the
other hand, a K-type population with small Balmer equivalent widths
(EWs) would lead to over-subtraction of the blue stellar continuum. In
the following, we will evaluate this uncertainty on a case by case
basis.

Comparison of emission line and stellar continuum fluxes shows good
agreement between the WFPC2 images and the spectral data (see
Table~2).  The 500~second F656N images have two saturated pixels 
corresponding to the H$\alpha$ emission peak. The effect of the
saturation is discussed in section~4.2.1; in the present context, the
large apertures employed for the image-spectrum comparison limit the
effect of the saturation on the measured H$\alpha$ flux to the
1.3\%~level.

\section{The Morphology of the Ionized Gas and of the Stellar Continuum}

At the distance of 4.1~Mpc (Sandage et al. 1994), the WF pixel size of
0\farcs1 corresponds to a spatial scale of 2~pc. Details over the
typical scale of starforming regions, $\sim$10~pc, can therefore be
studied with our images. 

Over scales of tens of arcseconds, the H$\alpha$ line emission is
roughly circularly symmetric, with many loops and filaments extending
mostly radially off the central region (see Figure~1a, and cf. Graham
1981, Martin \& Kennicutt 1995, Marlowe et al. 1995). Upon closer
inspection, the high spatial resolution of the WFPC2 image reveals a
wealth of complex structures in the ionized gas emission.  The E--W
dust lane which bisects the galaxy core along the minor axis is
visible in the H$\alpha$ map (Figure~1b). North of the lane, around
the high surface brightness region which identifies the nucleus, the
morphology of the gas emission is dominated by filamentary emission
and shell structures with scales of a few arcseconds. The peak of the
H$\alpha$ emission is roughly located at the center of the nuclear
region, and coincides with the position of a stellar cluster
(NGC5253-5, see Table~3). About 1\farcs3 NW of the H$\alpha$ peak, a
multiple shell structure with radii of about 20 pc each is visible in
Figure~1b. This is one of the positions where enhanced nitrogen
enrichment has been measured by Kobulnicky et al. (1997). The other
region of N enhancement found by the same authors is located about
1\farcs7 S of the center of the multiple shell structure. East of the
H$\alpha$ peak, a complex, roughly circular, structure of ionized gas
of about 65~pc diameter is present.  South of the dust lane, a number
of HII regions is distributed in the core amid a low level diffuse gas
emission. HII regions are a common feature across the entire H$\alpha$
image.  Two shell nebulae of about 25~pc diameter are located about
125~pc ENE and about 280~pc SSE of the core, respectively. The first shell,
and the brightest of the two, contains two blue stellar
objects, possibly ionizing stars. At the edges of the frame, at a
distance of about 0.7--0.8~kpc from the core, two extended filamentary
structures can be discerned ESE and W of the galaxy's center.

The profile of the H$\alpha$ surface brightness is shown in Figure~2
in annuli of increasing distance from the peak. The slope of the
H$\alpha$ surface brightness increases in absolute value from inside
out, and three different regions can be identified in the plot. The
inner region has a radius of about 3$^{\prime\prime}$, has the
flattest slope and has the highest surface brightness (the
nucleus). The intermediate region has a radius of about
13$^{\prime\prime}$ and includes the entire core of NGC5253.  Outside
the 13$^{\prime\prime}$ radius, the intensity of the H$\alpha$
emission drops rapidly. The surface brightness of the H$\alpha$
emission remains bright relative to the underlying stellar continuum
up to 20$^{\prime\prime}$ from the center; the radially averaged
EW(H$\alpha$) is above 90~\AA~ in the region. This implies that
intense photo- and shock-ionization of the ionized gas, and the star
formation, are active up to at least 400~pc from the core of the
galaxy. The presence of individual expanding bubbles, and enhanced
knots of emission, hundreds of parsecs from the core, support this
picture.

The morphology of the stellar continuum is markedly different from the
morphology of the ionized gas (see Figure~3; Martin \& Kennicutt
1995). The E--W dust lane which bisects the core is seen in the
stellar images as well, but the two regions North and South of the
lane do not show the dichotomy of appearance shown by the H$\alpha$
image. Indeed, the core appears quite homogeneous in terms of overall
structure and brightness. On small scales (a few arcseconds), the core
shows an irregular morphology, which alternates spots of high and low
stellar surface brightness, and is mostly due to the inhomogeneity of
the dust distribution.  The large scale ($>$15$^{\prime\prime}$)
elliptical isophotes of the stellar light profile are centered about
5\farcs8 SSW of the cluster NGC5253-5. The position of the starburst
nucleus is thus displaced relative to the center of the galaxy's
stellar light profile. 

\section{The Effects of Dust}

The dust reddening affecting nebular and stellar continumm emission in
the core of NGC5253 is analyzed in sections~4.1 and 4.2. In section
4.2, the specific cases of the stellar clusters and of the reddening
in the dust lane are discussed separately from the diffuse stellar
population (sections~4.2.1 and 4.2.2, respectively).  The reader
interested in a summary of the effects of dust in the galaxy can refer
to section~4.3.

\subsection{The Reddening of the Gas}

The H$\alpha$/H$\beta$ emission line ratio provides a measure of the
dust reddening affecting the ionized gas, since variations of the
intrinsic ratio are less than 5\%~ for a large range of electron
temperature and density. The H$\alpha$ and H$\beta$ emission line
images are thus used to construct a gas reddening map for NGC5253. The
images are first smoothed through median filtering with a
3$\times$3~pixel window to reduce noise fluctuations, and then
ratioed while setting to zero all pixels which are within 1.5~$\sigma$
of the noise level.  The result is shown in Figure~4a. The WFPC2
images reveal in detail the complex structure of the dust geometry,
where heavy obscuration alternates with almost reddening-free regions.
The E-W dust lane is the most apparent dust formation, together with a
filament which departs North of the lane and connects to the obscured
starburst nucleus. The region of highest obscuration is in the
starburst nucleus, with its peak about 0\farcs5 W of the H$\alpha$
emission peak. Two other elongated regions of large reddening are
located N-W and N-E of the nucleus.

The average value of the H$\alpha$/H$\beta$ ratio over the central
region of NGC5253 is 3.35, corresponding to A$_V$=0.53 for a
foreground homogeneous dust distribution. The extinction from our
Galaxy is A$_V$=0.16 (Burstein \& Heiles 1982), implying A$_V$=0.37
for the average intrinsic reddening. Thus, NGC5253 appears almost
reddening-free when {\it the average over an extended region} is
considered, despite the complex morphology of the line ratio map.  In
the nucleus and in the dust lane the observed H$\alpha$/H$\beta$ ratio
reaches values as large as $\sim$6 (values larger than 6 are dominated
by the noise), corresponding to optical depths of at least
A$_V\sim$2.2 assuming only foreground dust; those regions are
optically thick. 

The H$\alpha$/H$\beta$ ratio map provides information on the spatial
variations, integrated along the line of sight, of the reddening
affecting the gas in the galaxy. However, heavily obscured areas are
optically thick in both emission lines, and the H$\alpha$/H$\beta$
ratio does not constrain the total dust optical depth in these
regions. In the following section, we combine line and
continuum emission data to derive the reddening in mildly dusty
regions (A$_V\approx$1~mag). For the opaque nucleus, radio
observations from the literature are combined with our data to
constrain the extinction.

\subsection{The Reddening of the Stellar Continuum}

The small scale spatial variations of the stellar surface brightness
match very well the variations in reddening derived from the Balmer
emission lines (cf. Figure~3b with 4a). In addition, the regions with the
highest values of H$\alpha$/H$\beta$ correspond to the reddest regions
in the 547$-$814 color map (Figures~4a and 4b). All this indicates a
correlation between the dust obscuration of the stellar continuum and
of the gas emission. To quantify this statement, the behavior of the
colors 255$-$547 and 547$-$814 is analyzed as a function of
$\log(H\alpha/H\beta)$ for the central 16\arcsec$\times$16\arcsec~
(Figures~5a and 5b). Each point corresponds to the average color in a
0\farcs5$\times$0\farcs5~ bin; the binning has been chosen to ensure
that detections in each bandpass is at least 4~$\sigma$ above the
limit even in the most reddened, i.e., faintest, regions. The
uncertainties due to photon statistics for a 6~$\sigma$ detection, are
$\delta (255-547)$=0.11 and $\delta (547-814)$=0.07. Most of our data
are better than 6~$\sigma$ detections. The typical uncertainty in the
H$\alpha$/H$\beta$ ratio is of the order of 6\%, due to imperfect
knowledge of the correction for the underlying stellar
absorption. Both the Spearman and Kendall nonparametric correlation
tests show that the points in Figure~5a deviate 10.8~$\sigma$ and in
Figure~5b 11.9~$\sigma$, respectively, from the null correlation
hypothesis. An increase in the reddening of the gas is therefore
mirrored by a reddening of the mean stellar colors.

For a foreground, uniform dust distribution, the equation describing
the color-versus-line~ratio trend is a straight line defined as:
\begin{equation}
m_{\lambda_1}-m_{\lambda_2} = (m_{\lambda_1}-m_{\lambda_2})_0
+{\displaystyle k_{\lambda_1}-k_{\lambda_2}\over \displaystyle{0.4
(k_{\beta}-k_{\alpha})}}\log\biggl({R_{\alpha\beta}\over
R_{0,\alpha\beta}}\biggr),
\end{equation} 
where $(m_{\lambda_1}-m_{\lambda_2})_0$ is the intrinsic color of the
stellar population, $R_{\alpha\beta}$ and $R_{0,\alpha\beta}$ are the
observed and intrinsic H$\alpha$/H$\beta$ ratios, respectively, and
k$_{\lambda}$=A($\lambda$)/E(B$-$V) is the total-to-selective
extinction, calculated at the continuum wavelengths $\lambda_1$ and
$\lambda_2$ and at the wavelengths of the nebular emission lines
H$\beta$ and H$\alpha$. For the intrinsic H$\alpha$/H$\beta$ ratio in
NGC5253 we adopt the value 2.84 (Kobulnicky et al. 1997). In general,
the dust distribution will not be foreground and uniform, so we define
an {\it effective attenuation} $k'_{\lambda}$ and an {\it effective reddening} 
$k'_{\lambda_1}-k'_{\lambda_2}$ for Equation~(1).

The best fit of Equation~(1) to the points of Figures~5a and 5b (after
correction for the foreground Galactic reddening E(B$-$V)=0.05,
Burstein \& Heiles 1982) yields the values:
\begin{eqnarray}
(255-547)_0&=&-1.33\pm0.15\nonumber\\
(547-814)_0&=&-0.66\pm0.05,
\end{eqnarray}
and
\begin{eqnarray}
(k'_{255}-k'_{547})/(k'_{\beta}-k'_{\alpha})&=&1.20\pm0.14\nonumber\\
(k'_{547}-k'_{814})/(k'_{\beta}-k'_{\alpha})&=&0.52\pm0.04.
\end{eqnarray}
The intrinsic colors (255$-$547)$_0$ and (547$-$814)$_0$ are not
immediately comparable with the data at zero reddening in Figure~5
($\log(H\alpha/H\beta)\simeq$0.45); the large scatter in the data is
due to the presence of resolved stars and stellar clusters prevents
the comparison. A cross-check for Equation~(2) is obtained by
measuring the colors of the lowest obscuration regions in the core,
after removal of all identifiable clusters and stars.  With this
technique, the colors are 255$-$547=$-$1.1$\pm$0.1 and
547$-$814=$-$0.66$\pm$0.06, in remarkable agreement with the intrinsic
colors obtained from the best fit (Equation~(2)).

In order to understand the meaning of Equation~(3), we should remember
that, if the dust is located in a foreground homogeneous screen, the
selective extinction (A$_V$/E(B$-$V)$\simeq$3.1, Cardelli,
Clayton \& Mathis 1989) gives:
\begin{eqnarray}
\Bigl[(k_{255}-k_{547})/(k_{\beta}-k_{\alpha})\Bigr]_{ISM}&=&3.07\nonumber\\ 
\Bigl[(k_{547}-k_{814})/(k_{\beta}-k_{\alpha})\Bigr]_{ISM}&=&1.13.
\end{eqnarray} 

For diffuse ISM, this result is almost independent of the chosen
extinction curve, since in the wavelength range 2600-9000~\AA~ there
is little difference among the mean interstellar extinctions of the
Galaxy, the Large Magellanic Cloud, and the Small Magellanic Cloud
(Seaton 1979, Fitzpatrick 1986, Bouchet et al. 1985). The comparison
between Equation~(3) and (4) shows that the ratio of the effective
reddening of the stellar continuum to the effective reddening of the
gas emission in NGC5253 is smaller, by more than a factor two, than
the expected ratio in the case of foreground homogeneous dust. To
understand whether this discrepancy is due to the
stellar continuum or to the gas, it should be remembered that the
difference $k'_{\beta}-k'_{\alpha}$ is maximum if the dust is
foreground to the gas,
$$k'_{\beta}-k'_{\alpha}=k_{\beta}-k_{\alpha}=1.16;$$
more complex geometries than the simple foreground screen or dust
scattering into the line of sight will make the difference smaller
(CKS94) and will exacerbate the discrepancy between Equation~(3) and
(4). In addition, the small value of the ratio
$(k'_{547}-k'_{814})/(k'_{\beta}-k'_{\alpha})$ cannot be reproduced if
the geometrical distribution of the dust is the same for gas and stars
(CKS94, Calzetti, Kinney \& Storchi-Bergmann
1996). $(k'_{547}-k'_{814})$ probes larger optical depths than
$(k'_{\beta}-k'_{\alpha})$; however complex the dust distribution, if
gas and stars ``see'' the same amount of dust, the ratio between the
two quantities should be
$(k'_{547}-k'_{814})/(k'_{\beta}-k'_{\alpha})\ge 1.13$. Therefore, the
discrepancy between Equations~(3) and (4) implies that the dust
affecting the stellar emission has a covering factor about one half
that of the dust affecting the gas emission: the gas is associated
with regions of higher dust content than the stars.  One immediate
consequence of this scenario is that emission lines and stellar
continuum are subject to differential reddening, and the line EWs are
no longer reddening-free quantities (CKS94, Calzetti 1997). 

The results presented in this section should be intended as `mean'
reddening corrections for the stellar continuum in NGC5253. The large
spread in the data points of Figure~5 does not allow us to consider
such corrections as the `true corrections' for each individual data
point; rather, the corrections should be considered valid in a
statistical sense. 

\subsubsection{Reddening of The Stellar Clusters}

The reddening of the stellar clusters in the core of NGC5253 is 
analyzed on an individual basis, to aid the derivation of the clusters' 
ages in section~5.

Stellar clusters (see Table~3) generally populate areas of low dust
reddening in the core of NGC5253, with the (possibly only) exception
of NGC5253-5 (Figures~3b and 4a). The fact that clusters crowd in low
dust extinction regions may be an observational bias, since heavily
reddened clusters are more likely to go unnoticed. Alternatively,
evolving stellar clusters may destroy or `blow away' the surrounding
dust, via massive star winds and supernova explosions. This will make
the dust distribution inhomogeneous and create `holes' in the galaxy's
ISM, through which the clusters can be observed. Table~3 reports the
astrometric and photometric properties of the six brightest stellar
clusters in the core. The STMAG magnitude system will be used
throughout the paper. Two of the clusters (NGC5253-4 and NGC5253-5)
are in the starburst nucleus, North of the dust lane, while the other
four are located South of the lane. The photometry of the clusters was
carried out using a circular aperture of 0\farcs5 radius, and
subtracting the underlying stellar flux derived from a ring of
0\farcs5 size. Increasing the inner radius of the sky annulus and/or
the aperture size by a factor 2 introduces variations of $\delta
(255-547)\simeq\pm$0.07 and $\delta (547-814)\simeq\pm$0.06 in the
colors, and $\simeq$15--30\%~ in the fluxes. We adopt the variations
in the colors as uncertainties; we do not attempt to correct absolute
quantities for the aperture effects, but we bear in mind the presence
of this effect. Since the clusters are all located within the central
130~pixels of the frame, geometrical distortion and CTE effects are
negligible relative to the aperture effects (Holtzman et
al. 1995). The clusters are all resolved in the WF images, with
measured FWHMs in the range 0\farcs22--0\farcs37 (stars in the same
field have typical FWHMs in the range 0\farcs13--0\farcs17). The
deconvolved FWHMs correspond to physical sizes between 3.2~pc and
6.8~pc.

The emission line ratios indicate that corrections of the nebular
emission for the underlying stellar absorption further reduce the
already small reddening values of the clusters in Table~3. An
underlying absorption of EW=2~\AA~ at H$\alpha$ and H$\beta$ (see
McCall et al. 1985) has been assumed for all clusters. This correction
has negligible impact on the large EW of the line emission of the
clusters NGC5253-4 and NGC5253-5, for which a measurable residual
reddening is present. To bracket a range of possibilities, the
photometric quantities of the two clusters are corrected for dust
obscuration assuming two dust scenarios: 1) the small stellar
continuum reddening implied by Equation~(3) applies, thus attributing
to the clusters the same mean `effective reddening' observed for the
diffuse population; 2) the dust is entirely foreground to the cluster
and Equation~(4) applies. In the first case, the total-to-selective
obscuration value derived by Calzetti (1997) is adopted. A foreground
dust distribution ($k'_{\beta}-k'_{\alpha}=1.16$) is assumed for the
reddening of the gas. The difference in the final photometry given by
the different scenarios provide a measure of the uncertainties
introduced by the dust effects.

The very blue cluster NGC5253-4 `sits' at the edge of the obscured
nucleus (see Figures~3b and 4a). Despite its location, the emission
from NGC5253-4 suffers relatively low dust reddening; the cluster is
either situated on the foreground surface of the otherwise obscured
nucleus, or its massive star winds have succeeded in removing the
blanket of dust in front of the cluster during the past few Myr. We favor
the stellar winds mechanism, because the entire region surrounding the
cluster is characterized by low obscuration.  Uncertainties in the
reddening corrections have small effect on the age determination of
this cluster.

NGC5253-5 is the youngest cluster in the galaxy, with an age around
2.5~Myr (see section~5.2.1). The cluster is located roughly at the
center of the starburst nucleus, in a region characterized by some of
the largest values of the ratio H$\alpha$/H$\beta$.  The central two
pixels of the H$\alpha$ emission from the cluster are saturated; the
correction to the saturation has been determined assuming that the
H$\alpha$/H$\beta$ ratio of the saturated pixels is the same as the
ratio of the neighbouring unsaturated pixels.  The correction for the
H$\alpha$ flux is 26\% in the 0\farcs5 aperture, if the dust is
foreground; this is a lower limit, since more complex distributions of
dust will yield larger corrections for the H$\alpha$.  The observed
colors (Table~3) are much redder than the values 255$-$547=$-$2.5 and
547$-$814=$-$1.1 expected for a 2--3~Myr old cluster, and the
correction for reddening according to the two scenarios above is not
sufficient to account for the discrepancy. Possibly the two models
give inadequate reddening corrections for this cluster. The
alternative possibility, that NGC5253-5 harbors an AGN, has been
excluded by analyses of the high excitation lines (e.g., Lutz et
al. 1996); in addition, the observed colors of this source are too
blue for it to be an AGN (cf. Francis et al. 1991). 

Hence, the first hypothesis is likely to be true: the adopted models
for the reddening are inadequate. Indeed, such a young cluster would
be still embedded in the molecular cloud from which it originated. The
light from stars mixed with dust is subject to almost gray extinction
at large optical depths $\tau$, since the flux reduction due to
obscuration becomes proportional to 1/$\tau$: the spectral energy
distribution of the emitting object will appear almost unreddened even
in the presence of large amounts of dust (Natta \& Panagia 1984,
CKS94). A homogeneously mixed distribution of dust, stars and gas is
thus an efficient way to `hide' dust. In particular, the Balmer ratio
is no longer a good diagnostic of extinction: the maximum value for
the reddening between H$\alpha$ and H$\beta$,
A$_{\beta\alpha}=$A$_{\beta}-$A$_{\alpha}$, is 0.25~mag, for a dust
cloud with A$_V>$9~mag.  We measure A$_{\beta\alpha}$=0.35~mag in
front of NGC5253-5, implying that a small amount of dust
(A$_V\sim$0.3~mag) must be foreground to the cluster to account for
the additional A$_{\beta\alpha}$=(0.35$-$0.25)=0.1~mag.  The following
dust geometry in and around NGC5253-5 can account for the observed
colors: the cluster is behind a relatively thin dust layer with
A$_V\sim$0.3~mag, and is completely embedded in a thick dust cloud
with A$_V>$9~mag.  The colors and fluxes obtained from this dust model
are shown in Table~3, where we consider also the case that the thin
dust layer has optical depth A$_V$=0.45, the maximum value allowed by
the uncertainties.  Even after the new correction for reddening, the
547$-$814 color remains quite red relative to expectations for
clusters of a few~Myr age, while the 255$-$547 color is in close
agreement with expectations; we cannot exclude a contribution from red
stars foreground to the cluster.

Radio observations can help constraint the total obscuration in the
direction of NGC5253-5, since dust has little or no effect on radio 
emission. The position of the cluster is 1\farcs3 East and 0\farcs7
South of the 2~cm emission peak detected by Beck et al. (1996). The
2~cm emission from the nucleus of the galaxy has been shown to be 95\%
due to free-free emission from the ionized gas, so the ratio between
the H$\beta$ flux of NGC5253-5 and the radio emission can be used to
derive the total extinction towards the cluster. The ratio of the
emissivities between H$\beta$ and the radio is
j$_{\beta}$/j$_{2
cm}$=3.738$\times$10$^{-10}$~erg~s$^{-1}$cm$^{-2}$Jy$^{-1}$ (Caplan \&
Deharveng 1986) for a nebula with electron temperature
$T_e$=10$^4$~K. The flux at 2~cm is 23~mJy in the central radio source
(Beck et al. 1996), implying a flux of
8.6$\times$10$^{-12}$~erg~s$^{-1}$cm$^{-2}$ for H$\beta$. The observed
H$\beta$ flux (after correction for Galactic extinction) is
1.40$\times$10$^{-13}$~erg~s$^{-1}$cm$^{-2}$. To recover the large
flux implied by the radio emission, the stellar cluster should be
embedded in a dust cloud with a total optical depth of 35~mag, for a
homogeneous mixture of dust and stars. Again, a purely foreground dust
distribution is ruled out by the observed colors of the cluster.  We
have implicitly assumed that the position of the optical peak is
coincident with the radio peak, and that the observed difference of
1\farcs5 is due to astrometric uncertainties. We cannot entirely
exclude that the displacement between the two peaks is real and the
radio peak emerges from a region of higher dust obscuration than the
optical peak.  Also the FWHM of the radio peak, about 0\farcs8 (Beck
et al. 1996), is larger than the FWHM=0\farcs35 of the ionized gas
peak. The conclusion is that NGC5253-5 is embedded in a dust cloud
of optical depth between 9 and 35~mag at V (cf. Aitken et al. 1982).

\subsubsection{The Dust Lane}

The E--W dust lane suggests a dust geometry that is relatively easy to
model -- a sheet of dust (albeit inhomogeneous) embedded in a
distribution of stars. To test this geometry, we extract photometry on
either side of the lane and within it.  We assume that the diffuse
stellar population doesn't drastically change as a function of
location in and around the lane, except for a decreasing intrinsic
surface brightness as a function of the distance from the galaxy
center.  Additional constraints are added by the EW(H$\alpha$); areas
as close as possible in EW(H$\alpha$) are selected within and outside
the dust lane, to minimize differences in the stellar
population. Eight areas, three within the dust lane, and five adjacent
the lane, are identified, with 140~\AA$<$EW(H$\alpha)<$210~\AA.  Each
of the eight regions is a square of 15 pixels on the side; the size is
selected large enough to be roughly insensitive to local variations in
the diffuse stellar population; for the same reason, the regions
outside the dust lane are chosen as close as possible to the lane
itself, mantaining that they must be external to the area of high
reddening. The colors of these regions become redder for increasing
values of the observed H$\alpha$/H$\beta$ (Figure~6). We interpret
this result as an effect of dust reddening, rather than a change in
the intrinsic stellar population, for the following reason. If the
eight areas are radiation bounded, the observed values of
EW(H$\alpha$) imply an age range 150--500~Myr and a variation in
intrinsic colors $\delta$(255$-$547)$<$0.33 and
$\delta$(547$-$814)$<$0.03 for continuous star formation (Leitherer \&
Heckman 1995, LH95 hereafter). The observed range of colors in
Figure~6 is larger than what expected from age variations in the
stellar population.

Equation(1) is shown in Figure~6 overlaid on top of the data, using
both Equations~(3) and (4) for the selective reddening.  Equation~(3)
is in better agreement with the data than the simple foreground dust
geometry, implying, again, that the stellar continuum is on average
less reddened than the gas emission. The dereddened colors for these
regions are (255$-$547)$_0\simeq -$1.4 and (547$-$814)$_0\simeq -$0.7,
corresponding to stellar population undergoing constant star formation
over the past 200~Myr (Bruzual \& Charlot 1995, BC95 hereafter), in
agreement with the age range derived from the reddening-corrected
values of EW(H$\alpha$). This check ensures that Equation~(3) gives a
reasonable correction for the effects of dust extinction in the
region.

For our geometrical model, we assume the dust lane corresponds to a
clumpy layer of dust; regions of higher reddening correspond to a
larger number of clumps along the line of sight. We allow a fraction
f$_s$ of stars to be foreground to the lane, therefore `sandwiching'
it. The fraction f$_g$ of foreground gas is implicitly set to zero in
this model, because the quantity log(H$\alpha$/H$\beta$) is assumed to
scale linearly with the reddening. The dust clumps are assumed to be
all equal and Poissonian distributed (Natta \& Panagia 1984, see also
CKS94). The optical depth of the clumps at different wavelengths is
described by the SMC extinction curve (Gordon, Calzetti \& Witt 1997);
we also allow for the presence of scattered light into the line of
sight. Free parameters in this `sandwich' dust model are: the optical
depth of each clump and the fraction f$_s$ of foreground stars.  The
intrinsic colors of the stellar population are assumed to be in the
range $-$1.6$<$(255$-$547)$_0 < -$1.0 and $-$0.80$<$(547$-$814)$_0 <
-$0.60.  The best fit model (shown in Figure~6) is given for
A$_V$(clump)=0.16$\pm$0.02, f$_s$=0.10$\pm$0.01, and zero scattered
light into the line of sight from external regions. The dust lane
corresponds to an optical depth A$_V$(lane)$\simeq$2.2.  10\% of the
stars are in front of both the dust and the gas; therefore the stars
contributing to the UV and optical continuum emission must have a
scale length larger than the ionized gas and must `sandwich' the dust
lane to acccount for their reddening properties.

\subsection{Discussion: Dust in NGC5253}

Corrections for the effects of dust extinction are crucial even in the
case of a metal-poor dwarf galaxy like NGC5253. The inhomogeneity of
the dust distribution is the key to intepret the many observational
facts about the galaxy.  Despite the obvious presence of dust, large
aperture spectroscopy usually measures small reddening values
(CKS94). This behavior can be attributed partly to the mixed
dust/star/gas geometry of the nucleus, which produces almost gray
extinction, and partly to the intrinsically low reddening observed in
most of the galaxy's core.

Overall, the diffuse stellar population in the core is affected by a
smaller reddening than the ionized gas. This discrepancy in
obscuration between gas and stars is a common property of starburst
galaxies (CKS94, Fanelli, O'Connell \& Thuan 1988). The observed
properties are accounted for if the dust has a larger covering factor
in front of the ionized gas than in front of the stars. In the same
vein, the scale length of stars may be larger than the scale length of
the dust and of the gas emission. The stars located in
low reddening regions do not contribute significantly to the
ionization of the gas, and/or are not associated with ionized gas,
although they still contribute to the UV emission. In the first
scenario, holes in the ISM may have been created by the massive star
winds and supernova explosions of ageing stellar populations, and
presently older stars ``shine'' through these holes. In the second
scenario, nonionizing stars, which have longer lifetimes than ionizing
stars, diffuse across the galaxy. Stars have typical dispersion
velocities of 10~km\,s$^{-1}$ in our Galaxy, and the brightest O stars
have a lifetime of approximately 3~Myr: they would travel only 30~pc
from their place of birth over their life, while less massive stars
would travel longer distances. In NGC5253, the half-light radius of
the H$\beta$ emission in the WF3 is R$_e$=6\arcsec, while it is
R$_e$=12\farcs5 for the UV stellar emission. The ionized gas emission
in this galaxy is thus more concentrated than the blue stellar light,
and emerges from the heavily obscured starburst nucleus, while the
UV-emitting stars are distributed across the galaxy's starforming
core.  The two scenarios, holes in the ISM and stellar diffusion, are
not mutually exclusive, and both can account for the observed behavior
of the stellar reddening. CKS94 derived an effective obscuration curve
for the stellar continuum in starbursts (see, also, Calzetti 1997,
Gordon et al. 1997) which folds together dust extinction
and geometrical effects; this curve reproduces well the values of the
selective reddening reported in Equation~(3) and can therefore be used
to describe the reddening characteristics of the stellar continuum in
NGC5253.

Young starburst regions tend to be obscured: here the massive star
winds and the supernova explosions have not had time to blow away the
dust of the parental cloud in which the stars are embedded. In the
nucleus of NGC5253, the almost complete absence of non-thermal radio
emission (e.g., Beck et al. 1996) confirms that there have not been
yet enough supernova explosions to help this process. The dust
geometry is therefore dominated by the presence of the
molecular clouds.  The reddening in the nucleus of NGC5253 is
generally high, although largely variable, with optical depths at V
which range from 0.4~mag to 10~mag, and possibly higher at the 
position of the cluster NGC5253-5, in agreement with results
obtained from infrared data (e.g., Aitken 1982, Telesco et
al. 1993). Models of foreground dust are not sufficient to account for
the obscuration, and the dust needs to be mixed with the stars and the
gas. Because of the proximity of the dust to the ionizing stars, the
former can be efficiently heated to high temperatures, explaining the
higher mid-infrared to far-infrared ratio of NGC5253 relative to other
starburst galaxies (Roche et al. 1991, Telesco et al. 1993, see, also,
Calzetti et al.  1995). A complete description of the complex geometry
of the nucleus is not possible with the data in hand, due to the
limited range of optical depth sampled by the UV and optical
data. Such a study would be immensely helped by high spatial
resolution infrared mapping, as possible with the NICMOS instrument
onboard HST.

\section{The Ages of the Stellar Populations}

The results from section~4 are used here to break the age/reddening
degeneracy and derive ages for the diffuse stellar population
(section~5.1) and for the stellar clusters (section~5.2). The age
information is used in section~5.3 to derive a picture of the recent
star formation history in the core of NGC5253.

\subsection{The Diffuse Stellar Population}

Constraints on the age of the diffuse stellar population can be
derived from color-color diagrams and from the EW of the emission
lines. Figure~7 shows the colors of the galaxy's
mean stellar population integrated along the line of sight for the
central 16\arcsec$\times$16\arcsec, corresponding to a physical region
of 320$\times$320 pc$^2$.  The mean value of the colors is not
dominated by the contribution from resolved stars, as it does not
depend on the size of the bins selected to calculate the colors
themselves; however, the spread in the data is probably dominated by
such effect (see the discussion in section~4.2). The contribution of
the sky emission to the 547 and 814 bands is about 5\%~ and 6\%~,
respectively, and is negligible in the 255 band. To correct the colors
for the effects of dust reddening, Equations~(1) and (3) are applied to
the data points. The EW(H$\alpha$) is corrected for differential
reddening between gas emission and stellar continuum using the results
from section~4.2 and the total-to-selective reddening of Calzetti
(1997): the difference in effective reddening between gas and stars is
k$^g_{H\alpha}-$k$^s_{H\alpha}\simeq$0.48.  The reddening-corrected
colors and EW are reported in Figures~8a--c.  The comparison between
Figure~7 and 8a shows the importance of reddening correction even in
the case of a metal-poor galaxy like NGC5253: the 255$-$547 color
becomes bluer by about $-$0.5~mag in mean value after the effects of
dust are removed, leading to a factor 2 smaller inferred age for the 
stellar population.

Synthetic models of stellar populations must be employed to connect
colors and EWs to ages. The population synthesis models of BC95 are
plotted in Figure~8a for different age values of a stellar population
undergoing constant star formation with a Salpeter ($\alpha$=2.35) IMF
in the stellar mass range 0.1--100~M$_{\odot}$. Although NGC5253 has
metallicity 1/5--1/6~solar (e.g., Walsh \& Roy 1989), the solar
abundance BC95 models are adopted here; models with sub-solar
abundances are still inadequate to account for the colors of red
supergiants, and therefore the optical--IR colors of evolving
populations (Goldader, Leitherer \& Schaerer 1997). The contribution
of the nebular continuum, taken from LH95, has been added to the BC95
stellar population spectra, before producing synthetic colors in STMAG
via the STSDAS task SYNPHOT. Uncertainties on the photometric
zero-point are included in the error-bar shown in Figure~8a. The
variations of EW(H$\alpha$) for constant star formation as a function
of age are from LH95 (Figures~8b--c).  The data cover the age range
between 5$\times$10$^7$~yr and 1~Gyr, with the peak of the
distribution in the interval 100--500~Myr. This age range should be
considered an {\it effective} age, meaning that different combinations
of stellar populations can produce the same colors and EW(H$\alpha$).
The colors of the regions with the highest values of EW(H$\alpha$)
appear undercorrected for reddening (see Figures 8b--c). Such high
values of the EW are likely to be associated with very young stellar
populations; for these regions, the dust obscuration may be more
complex than described by Equations~(1) and (3)); furthermore, the
continuous star formation assumption may be inadequate. Finally,
EW(H$\alpha$) may not be a good age discriminant if the gas is ionized
by stars other than the underlying ones, or if the emission is
non-thermal.

There are reasons to believe that the nucleus contains a stellar
population which is younger than the population in the core as a
whole; the detection of W-R stars and the mean
EW(H$\alpha$)$\sim$1000~\AA~ indicate the presence of a starforming
population with an age of a few Myr in the nucleus. This young stellar
population is not entirely identifiable with the two nuclear stellar
clusters, NGC5253-4 and NGC5253-5, since the total H$\alpha$ flux and
the total number of W-R stars are larger than what can be accounted
for by the two clusters (see section~5.2.1). Unlike the
case of the core population, the reddening-corrected color-color
diagram of the nuclear region is not in agreement with either models
of constant star formation or of instantaneous burst populations
(Figure~9). The color 547$-$814 is the most deviant from the models;
this apparent ``undercorrection'' of the red bandpasses relative to
the blue bandpasses is typical of when foreground reddening
corrections are applied to regions where the dust is actually mixed
with the stars (cf. Natta \& Panagia 1984, CKS94).  From the 255$-$547
colors of the starburst nucleus, the age limits are $<$100~Myr for
continuous star formation, and $<$20~Myr for an instantaneous burst
population. More stringent age limits can be placed from the EW of the
nebular lines: 95\% of the data points have EW(H$\alpha$)$>$250~\AA~
and EW(H$\beta$)$>$65~\AA, implying ages $<$20~Myr for constant star
formation and $<$5~Myr for an instantaneous burst. The peaks of the
distributions are at EW(H$\alpha$)=1400~\AA~ and EW(H$\beta$)=280~\AA~
consistent with ages of $\sim$5~Myr and 3~Myr for constant star
formation and instantaneous burst, respectively (LH95). As already
mentioned, the ionized gas is probably not co-spatial with the ionizing
stars; we expect that the EW of the Balmer emission lines place an
upper limit to the age of the starburst nucleus.

\subsection{The Stellar Clusters}

We assume that each stellar cluster can be described by a co-eval
stellar population and use instantaneous burst population models to
estimate its age from the colors and the EW of the H$\alpha$ and
H$\beta$ lines. The latter are treated only as upper limits to the
ages, since our aperture (equivalent to about 10~pc) may not include
all of the emission line gas ionized by the stellar cluster.
 
Models of instantaneous burst populations are constructed from both
empirical stellar libraries (BC95) and theoretical stellar libraries
(LH95 and Bruzual \& Charlot 1996, BC96 hereafter) using solar
metallicity and Salpeter IMF in the range 0.1--100~M$_{\odot}$.
Contribution of the nebular continuum to the colors from LH95 is added
to all the models. The two types of libraries differ in the predicted
colors. The 547$-$814 color predicted by BC95 is systematically redder
by 0.2--0.3~mag than that predicted by LH95 and BC96 for ages
$>$7--8~Myr. The four clusters older than 8~Myr in our sample (see
Table~3) have 547$-$814 in better agreement with BC95 than with the
other models. BC95 predicts 255$-$547 colors which are about 0.3~mag
redder than LH95 and BC96. This effect is attributed to inadequate
correction for reddening of the massive stars in the spectral library
of BC95, which produces a ~30\% depression of the UV flux at 2600~\AA~
in the stellar population models. The uncertainties from the models,
together with the uncertainties from the reddening corrections, are
factored in the final age value of each cluster. Age estimates on the
basis of the EW of H$\alpha$ and H$\beta$ are done using the LH95
models. The derived age ranges are listed in the last column
of Table~3.

\subsubsection{Clusters in the Starburst Nucleus.}

The cluster NGC5253-5, which resides in the center of the starburst
nucleus, is the youngest cluster in the galaxy: the EW of its H$\beta$
emission indicates an age of 2.2~Myr, while the EW of H$\alpha$
indicates an upper limit to the age around 2.8~Myr. The FWHM of the
H$\alpha$ emission at the position of the cluster is comparable with
the FWHM of the stellar emission from the cluster, suggesting that the
ionized gas is as concentrated as the stellar cluster and, likely,
co-spatial.  The total H$\alpha$ emission within an aperture of
0\farcs5 radius centered on NGC5253-5 is
5.84$\times$10$^{-13}$~erg~s$^{-1}$cm$^{-2}$, with an uncertainty
around 10\%, due mostly to uncertainty in the saturation
correction. Of this flux, about 80\%~ comes from the cluster proper
(cf. Table~3); the remaining 20\%~ can be attributed to
photoionization from the diffusely distributed stars in the nuclear
region (as inferred from the H$\alpha$ intensity {\it surrounding the
cluster}). As seen in section~4.2.1, the colors of NGC5253-5 cannot be
used as effectively as the EW of the emission lines to derive an age
because of the heavy reddening in the direction of this cluster.  The
detection of a Wolf-Rayet feature in the spectrum of NGC5253-5
(Schaerer et al. 1997) indicates an age of at least 3~Myr, a slightly
greater value than derived from the nebular lines . The feature detected
by Schaerer and collaborators may be due to very massive
(M$\sim$100~M$_{\odot}$) stars with strong mass loss which mimic
Wolf-Rayet stars; these stars are 2~Myr old at most, and are thus
still on the main sequence (de Koter, Heap, \& Hubeny 1997).  In
addition, the cluster is located on top of a generally active
starforming area a few arcseconds in size which may include W-R
stars, a fraction of which would be encompassed by the slit aperture
used by Schaerer et al.

If the dust cloud in which NGC5253-5 is embedded has optical depth
A$_V$=35~mag, the intrinsic optical magnitude of the cluster is
m$_{547}$=13.5, corresponding to M$_V$=$-$14.5, and a mass in stars of
about 10$^6$~M$_{\odot}$, for the adopted IMF.  On the basis of its
absolute magnitude and its half-light radius, R$_e\sim$3.5~pc,
NGC5253-5 can be classified as a `super-star-cluster', comparing well
to the super-star-clusters in other blue compact dwarf galaxies
NGC1705 (Meurer et al. 1992), NGC1569 (O'Connell, Gallagher \& Hunter
1994), He2-10 (Conti \& Vacca 1994), as well as in other starburst
galaxies (Hunter, O'Connell \& Gallagher 1994, Watson et al. 1996).
Even assuming that there are ``only'' 9~mag of optical extinction
towards NGC5253-5, the rate of ionizing photons, 1$\times$10$^{52}$
photons/s, and the intrinsic optical magnitude, M$_V$=$-$13,
correspond to a mass in stars of about 2--3$\times$10$^5$~M$_{\odot}$:
NGC5253-5 thus remains a massive cluster.

NGC5253-4 is the brightest and has the bluest observed colors among
the UV stellar clusters in the core of NGC5253 (Meurer et al. 1995). A
young age and the low dust reddening are implied by the observed
characteristics. The total H$\alpha$ flux within the 0\farcs5 radius
aperture is 1.70$\times$10$^{-13}$, comparable to the flux measured by
Kobulnicky et al. (1997) from their FOS spectrum in the 0\farcs9
diameter aperture; the background-subtracted H$\alpha$ flux is only
70\%~ of that value (Table~3). As in the case of NGC5253-5 the
remaining 30\%~ of the H$\alpha$ flux can be attributed to the
underlying starburst population. The colors indicate an age around
2.5~Myr, while the EW of the hydrogen recombination lines indicate an
upper limit to the age between 4.0 and 4.4~Myr.  The properties of
NGC5253-4 are well explained by a moderately aged instantaneous burst
of star formation. Our derived age is confirmed by the detection of
Wolf-Rayet stars in this cluster, which constrains the age to be
around 3--4~Myr (Schaerer et al. 1997). The number of ionizing photons
currently produced by the cluster is $\sim$2.5$\times$10$^{50}$
photons/s, corresponding to a total number of about 25 O stars
(LH95). The typical WR/O ratio is between 1/3 and 1/10, implying that
at most 5--8 WR stars should be present in NGC5253-4 (Vacca \& Conti
1992, LH95). Of the 42 WR stars detected by Schaerer et al. in their
1\farcs6 long slit aperture spectrum, only about 1/5 are thus
associated with the stellar cluster, while the others are probably
associated with the diffuse starburst population. Indeed, NGC5253-4 is
surrounded by a number of resolved blue stars which appear to be
associated with the cluster, though they may not be gravitationally
bound to it. The total mass in stars of NGC5253-4 is about
1.3$\times$10$^4$~M$_{\odot}$. A similar number for the mass in stars,
4$\times$10$^4$~M$_{\odot}$, is obtained from the F547M magnitude
(LH95). NGC5253-4 is therefore a relatively small cluster, even for an
IMF which extends down to 0.1~M$_{\odot}$. Its absolute magnitude,
M$_V\simeq$M$_{547}$=$-$11, is comparable to the absolute magnitude of
R136 in 30~Doradus (e.g., Leitherer 1997), and the cluster has
half-light radius R$_e$=3~pc, slightly smaller than the radius of
NGC5253-5.

The youth of NGC5253-4 and NGC5253-5 is consistent with the age range 
derived for the diffuse stellar population in the nucleus (see
previous section) and the almost complete absence of non-thermal radio
emission in the center of NGC5253 (e.g., Beck et al. 1996): the first
supernovae are expected to explode after 3--4~Myr (LH95), and the
region has not yet or has just reached that stage.

The ratio of the 2~cm emission from the compact source to the 2~cm
emission from the entire nucleus is about 2/3 (Beck et al. 1996). If
the ratio of ionizing photons from the clusters and from the nucleus
is approximately the same, 65\%~ of the nuclear star formation is
happening in the two clusters. At minimum, 20\%~ of the nuclear star
formation is happening in the two clusters, as inferred from the ratio
of the H$\alpha$ flux from the clusters to the total H$\alpha$
emission from the region
(5.71$\times$10$^{-12}$~erg\,s$^{-1}$cm$^{-2}$, after correction for
extinction). For comparison, the stellar clusters in the entire
starforming core of NGC5253 represent at most 15\% of the UV
light. The contribution of cluster emission thus appears to be more
important in the nucleus than in the core as a whole; the difference
in age between nucleus and core suggests that cluster formation is an
important mode of star formation in the early phases of a starburst
event.

In order to derive ages for the stellar clusters and for the diffuse
stellar population, the stellar IMF has been assumed {\it a priori} to
be Salpeter-like ($\alpha$=2.35) with mass range
M=0.1--100~M$_{\odot}$.  Some {\it a posteriori} considerations, based
on the value of the EW(H$\beta$) of NGC5253-5, suggest that the
assumption on the slope and the upper mass limit may be not too far
from the reality.  An upper mass limit of only 30~M$_{\odot}$ or a
steeper IMF slope ($\alpha$=3.3) would produce an EW of H$\beta$ less 
than half that observed. This is indirect evidence that the
adopted values for the slope and the upper mass limit describe
adequately the observed properties in NGC5253, in agreement with
previous findings on dwarf galaxies (Massey et al. 1995, Stasi\'nska \&
Leitherer 1996), and recent suggestions on massive galaxies (Calzetti
1997, Bresolin \& Kennicutt 1997).

Note that some of the derived quantities, such as cluster masses and
star formation intensities (see previous and next sections), are more
sensitive to variations of the low-mass end of the IMF, for which we
have no direct information.  As an example, if the IMF low-mass
cut-off is at 1~M$_{\odot}$ rather than 0.1~M$_{\odot}$, all cluster
masses and star formation intensities decrease by a factor 2.5.

\subsubsection{Older Clusters in the Core}

The other four of the stellar clusters of Table~3 are older than
NGC5253-5 and NGC5253-4, with ages between 10 and 50~Myr, as
determined from the colors. Their absolute magnitudes are in the range
M$_V$=$-$10 to $-$11.3; because of aging, they have lost between 1 and
2.5 magnitudes relative to their peak brightness (LH95). Their masses
are in the range 0.7--4$\times$10$^5$~M$_{\odot}$.  These clusters are
comparable in mass to NGC5253-5, and are almost one order of magnitude
more massive than NGC5253-4.  Taken all together, the six clusters
have half-light radii in the range 1.6--3.5~pc, and ages in the range
2--50~Myr.  The oldest clusters tend to have the smallest radii: the
two clusters in the nucleus have half-light radii in the range 3--3.5~pc,
while the four other clusters have radii in the range 1.6--2.9~pc. The
cluster's crossing time for a star with velocity 5~km/s is less than
2~Myr; this implies that at least the four clusters older than 10~Myr
are likely to be gravitationally bound, a suggestion suppoerted by the 
decreasing radius for increasing age. 

\subsection{Discussion: Recent Star Formation}

The bulk of the current star formation in NGC5253 is located in an
area of radius $\sim$2\farcs5--3$^{\prime\prime}$, i.e. 50--60~pc, in
the North part of the core, centered at the position of the H$\alpha$
peak. The entire region appears not older than about 10~Myr, and is
probably as young as 5~Myr. The age of the starburst is suggested by a
series of clues: the region contains two very young clusters, with age
between 2 and 4~Myr; the H$\alpha$ surface brightness is high, with an
EW peaking above 1000~\AA; the radio emission is almost entirely
thermal (Beck et al. 1996); and the number of red supergiants is small
(Campbell \& Terlevich 1984). The two N-enriched regions in the
starburst nucleus are probably observed when massive stars are
polluting the medium (timescale of 10$^6$~yr), but before supernovae
explosions can homogenize the ISM (timescale of 10$^7$~yr, Kobulnicky
et al. 1997). The average star formation intensity in this area is
between 10$^{-5}$ and 10$^{-4}$~M$_{\odot}$/yr/pc$^2$ for a
0.1--100~M$_{\odot}$ Salpeter IMF, depending on the adopted value for
the reddening. The value corresponds to the maximum starburst
intensity found in galaxies (Meurer et al. 1997a).

In the part of the core South of the dust lane, the H$\alpha$ surface
brightness is smoothly distributed, except for a few isolated and
relatively faint HII regions/complexes; the corresponding star
formation rate is about 10$^{-6}$~M$_{\odot}$/yr/pc$^2$, one to two
orders of magnitude lower than in the nucleus. The ionized gas
emission is probably produced by the same diffuse population(s), which
is(are) responsible for the blue colors in the core of NGC5253. The
four bright clusters with ages between 10 and 50~Myr are situated also south
of the dust lane, within a region of about 10\arcsec~ in size, and
have no longer a dominant role in ionizing the gas. These clusters
appear to be the remnants of a past star formation episode in the
region.  A starburst event thus started about 50-100~Myr ago or
earlier and produced clusters which have the same characteristics
(luminosity, physical size) as the two young ones in the nucleus. Star
formation is still going on in the core, albeit at a lower efficiency
level than in the nucleus. The blue colors of the core are equivalent
to a stellar population which is continuously forming stars since
100--500~Myr, in agreement with the above picture. The core of NGC5253
thus appears to have undergone two or more starburst episodes over the
last $\sim$100~Myr or so, or a continuous star formation episode of
variable intensity.  The oldest stellar clusters may, however, not be
directly linked to the star formation history of the area where they
are located. With a typical velocity dispersion of 46~km\,s$^{-1}$
(Caldwell \& Phillips 1989), the crossing time of the core region
(about 200~pc) is less than 10~Myr.  The difference in age between the
diffuse stellar populations in the nucleus and in the core as a whole
explains the difference in the half-light radii between the UV stellar
emission and the ionized gas emission (R$_e$(UV)=12\farcs5 versus
R$_e$(H$\beta$)=6\arcsec) in NGC5253.  While only stars more masssive
than 20--30~M$_{\odot}$ produce hard enough radiation to ionize the
gas, stars with masses as low as 5--10~M$_{\odot}$ can significantly
contribute to the UV emission at 2600~\AA. Typical lifetimes of the
former are $\sim$10~Myr, while the latter can live as long as 100~Myr
or so.

The central starburst in NGC5253 is just the very last episode of a
complex star formation history, of which we have glimpsed the last
100~Myr through the analysis of the stellar clusters and of the
diffuse blue population. Burst durations have implications for the
detection of the post-starburst galaxies.  Norman (1991) calculated
that if bursts of star formation are almost instantaneous events, as
many starburst as post-starburst galaxies should be observed, after
making allowances for the dimming of the aged burst. The number of
observed post-starburst galaxies is much smaller than the number of
starbursts, implying either that starbursts do not form A-type and
lower mass stars or that bursts of star formation are long-lived
events.  Our results point in the direction of starbursts as
long-lived events, although we cannot exclude that low mass stars
do not form (cf. Calzetti 1997).  A burst duration of 100~Myr reduces
the post/burst ratio to about 1/2 (Norman 1991), and longer lifetimes
reduce the ratio even further.

\section{Summary}

The analysis of the HST WFPC2 images of NGC5253 has shown that 
the observational peculiarities of this metal-poor dwarf galaxy are
mostly driven by the combination of the dust distribution and of the
evolution of the star formation in the core.

The dust is inhomogeneously distributed across the central
20\arcsec$\times$20\arcsec, alternating regions of small and large
reddening. On a scale of a few arcsec, areas of high obscuration for
the gas are correlated with areas of high obscuration for the stellar
continuum. The highest values of obscuration are found in the E-W dust
lane and in the starburst nucleus. The former can be modeled by a
clumpy dust layer of optical depth A$_V\simeq$2.2, which is obscuring
the gas and most of the stars located in the region; a small fraction,
$\sim$10\%, of the stars is in front of the dust lane, accounting for
the smaller reddening of the stellar continuum relative to the ionized
gas.  In the nucleus, the configuration which explains the observed
stellar colors is a homogeneous mix of dust, stars, and gas; the stars
generated by the starburst are still embedded in the parental cloud,
which has a total optical depth at V of 9--35~magnitudes (cf. Aitken
1982). The rest of the core region is compatible with relatively small
values of the dust reddening, which however still imply corrections of
about $-$0.5 mag in the 255$-$547 colors. This corresponds to
a decrease in the age of the stellar populaiton of about a factor 2.

The simultaneous presence of an obscured starburst nucleus and an
almost dust-free blue core accounts for both the intense UV and
infrared emission.  The starburst nucleus provides the bulk of the
ionized gas emission: half of the H$\alpha$ flux from the galaxy is
within a radius of 6\arcsec~ from the peak of the emission. Mixed
configurations of dust and gas produce almost grey extinction,
accounting for the relatively small values of the reddening derived in the
nucleus from the Balmer decrement. The UV light, on the other hand,
comes from a larger region: half of the emission at 2600~\AA~ is from
a radius of 12\arcsec. The bulk of the observed UV light is thus
provided not by the young starburst nucleus, but by the core as a
whole, which is older, almost unextincted, and is producing stars at
an efficiency rate about 10 times lower than the nuclear
starburst. The dichotomy in the origin of the ionized gas and UV
stellar emission explains why the UV spectral energy distribution of
the core is not as blue as expected for a young, unextincted starburst
(Gonzalez-Riestra et al. 1987, CKS94). Despite the inhomogeneity of
the dust distribution, a general result can be derived for the
reddening correction in the center of NGC5253: the stellar continuum
suffers on average about half the reddening of the gas emission. This
behavior is typical of the nuclei of starburst galaxies, and is
explained if the gas is more closely associated with the dust than the
stars and much of the observed UV and optical continuum emission is
provided by non ionizing stars located in low extinction area (CKS94,
Calzetti 1997). The case study of NGC5253 has additionally
demonstrated that the dichotomy in reddening behavior cannot be
attributed to bulk diplacements between the stellar continuum and the
ionized gas emission; in fact, the dichotomy holds, on average, down
to scales of a few pc.

The six brightest stellar clusters in the core span an age range
between 2 and 50~Myr. Age is the largest discriminant among them;
other characteristics, such as absolute luminosities (scaled to a
coeval value), sizes, and inferred masses, have similar values. Five
of the clusters have masses around 10$^5$~M$_{\odot}$, for a Salpeter
IMF extending down to 0.1~M$_{\odot}$. The half-light radii range from
1.6 to 3.5~pc, with the older clusters being more compact than the
younger ones. The clusters are probably gravitationally bound. The two
youngest clusters, with ages between 2 and 4~Myr, produce between 20\%
and 65\% of the ionizing photons in the nucleus, a higher fraction
than produced by the clusters in the core, as expected if the early
stage star formation is associated with cluster formation. The other
four, older clusters (10-50~Myr) are located south of the starburst
and are roughly co-eval with the diffuse starforming population of the
core. There appears to be an anti-correlation between age and
reddening of the cluster, with the youngest object being more heavily
extincted than the oldest.  Stellar clusters are thus borne completely
enshrouded in the dust of the parental cloud, but after only 2--3~Myr,
emerge from the cloud and become UV bright. The similarity among the
six clusters in term of masses and sizes suggests that they were
generated by similar processes.

The age range of the clusters and the intrinsic colors of the diffuse
stellar population in the core suggest that star formation has been
fairly active in the region, albeit with variable intensity, during at
least the past 100~Myr. The current peak of star formation is more
spatially concentrated than the star formation integrated over the
last 100~Myr. The nuclear burst of star formation, with an average age
around 5~Myr, thus represents only the last of a sequence of starburst
episodes. The timescale of the star formation activity in the core may
ultimately be much longer than 100--500~Myr; this analysis of the
stellar population content of NGC5253 will be refined in a forthcoming
paper (Meurer et al. 1997b).

\acknowledgments

The authors would like to thank Robert Fesen for providing information
on the observing strategy adopted for the ultraviolet images of
NGC5253.  This research has made use of the NASA/IPAC Extragalactic
Database (NED) which is operated by the Jet Propulsion Laboratory,
California Institute of Technology, under contract with the National
Aeronautics and Space Administration.  During this work, G.R.M. and
D.R.G. were supported by the NASA Grant GO-06524.01-95A.

{}

\newpage
\figcaption[hst5253_fig1a.ps,hst5253_fig1b.ps]{a) (Plate) The
continuum-subtracted H$\alpha$ image; the scale is 52\arcsec~ on a
side. North is up; East is left. The image is shown in logarithmic
contrast to emphasyze the low surface brightness regions. b) (Plate) The inner
25\arcsec$\times$25\arcsec~ detail of the H$\alpha$ emission; the
contrast of the image is in linear scale.}

\figcaption[hst5253_fig2.ps]{The continuum-subtracted H$\alpha$
surface brightness as a function of the distance from the central
peak. The size of the 1~$\sigma$ uncertainty is shown as an orizontal
bar inside the circles of the data points.}

\figcaption[hst5253_fig3a.ps,hst5253_fig3b.ps]{a) (Plate) Three-color
image of the central 52\arcsec$\times$52\arcsec of NGC5253 in the
stellar continuum at 2600~\AA~ (F255W, blue), 5500~\AA~ (F547M, green)
and 8000~\AA~ (F814W, red).  North is up; East is left. The image has
the same size, orientation, and position of the H$\alpha$ image of
Figure~1a. b) (Plate) The F547M image of the inner
25\arcsec$\times$25\arcsec. The scale and position are the same as
Figure~1a. The six brightest stellar clusters are marked with the same
ordering given in Table~3.}

\figcaption[hst5253_fig4a.ps,hst5253_fig4b.ps]{a) (Plate) The
H$\alpha$/H$\beta$ ratio image for the inner
25\arcsec$\times$25\arcsec. North is up; East is left.  Light regions
indicate small values of the line ratio (low reddening), while darker
regions mark large values of the reddening. The white edges indicate
lack of signal.  Three spots 1--2 pixels wide of lighter shade can be
discerned inside the central region; the spots correspond to the
positions of stellar clusters, and the low value of H$\alpha$/H$\beta$
measured in those pixels is due to imperfect matching of the F487N and
F656N PSFs and slight mis-registration. b) (Plate) The 547$-$814 map
is shown on the same scale as the H$\alpha$/H$\beta$ image; regions of
high reddening in the H$\alpha$/H$\beta$ map correspond to red colors
(dark regions) in the stellar continuum map.}

\figcaption[hst5253_fig5a.ps,hst5253_fig5b.ps]{a) The 255$-$547 color
plotted as a function of the gas reddening indicator
H$\alpha$/H$\beta$ for the inner
16$^{\prime\prime}\times$16$^{\prime\prime}$ of the galaxy. Each data
point corresponds to a region 0\farcs5 in size. The typical
uncertainty on the data is reported at the top right of the plot. The
vertical bar to the left of the data points represents the colors of
the diffuse stellar population after removal of stellar clusters and
resolved stars. The best fit straight line is drawn through the data
points (dot-dashed line).  b) The same, for the 547-814 color.}

\figcaption[hst5253_fig6a.ps,hst5253_fig6b.ps]{a) and b) as in
Figure~5, now for the 8 regions selected inside and around the E--W
dust lane. The two straight lines represent: the foreground dust layer
(dotted line) and the `effective reddening' curve derived from
Equation~(3) (continuous line). The curved line represents the
`sandwich' dust model (dot-dashed line.}

\figcaption[hst5253_fig7.ps]{Color-Color plot of the inner
16$^{\prime\prime}\times$16\arcsec. Each data point corresponds to a
bin 0\farcs5 in size. The colors 255$-$547 and 547$-$814 are in
the STMAG system.  The straight line at the bottom left corner is the
reddening vector for A$_V$=1~mag. The error bar at the top right
corner shows the uncertainty given by the photon statistics of a
6~$\sigma$ detection, convolved with the photometric
uncertainties. The vast majority of the data points in the plot have
detection level higher than 6~$\sigma$.}

\figcaption[hst5253_fig8a.ps,hst5253_fig8b.ps,hst5253_fig8c.ps]{a) As
in Figure~7, after correction for dust reddening applying the observed
mean obscuration (Equation~(3) and (1)).  The error bar at the upper
right corner of the figure is as in Figure~5. The continuous line
marks the evolutionary sequence for the colors of a stellar population
with constant star formation, in the age range 1~Myr--15~Gyr; the
filled triangles mark a set of relevant ages in this sequence: 1, 3,
5, 8, 10, 20, 50, 100, 200, 500, 1000, 5000, and 15000 Myr.  b) and c)
The colors of the inner 16$^{\prime\prime}\times$16\farcs~ are plotted
as a function of the H$\alpha$ line equivalent width after reddening
correction (see text). The 1~$\sigma$ uncertainty is shown in the top
left corner of each diagram. The colors are the b) 255$-$547 and c)
547$-$814.  The continuous line marks the evolutionary sequence of a
stellar population with constant star formation, in the age range
1--500~Myr (LH95); the filled triangles 
have the same age separation as panel a).}

\figcaption[hst5253_fig9.ps]{Color-color plot of a
5\farcs5$\times$5\farcs5 region of the galaxy centered on the H$\alpha$
emission peak, at the position of the starburst nucleus. Each data
point is a 0\farcs5 square bin.  The surface brightness of each bin
has been galaxy-background subtracted; the background has been derived
from an annulus of 3\arcsec~ inner radius, and 2\arcsec~
thickness. Only points with fluxes at least 2~$\sigma$ above the
defined background have been reported in the plot. The data are
corrected for dust reddening following Equations~(1) and (3). The
continuous line corresponds to continuous star formation at the
different age values indicated in the figure. The dot-dash line
corresponds to an instantaneous burst population in the age range
1--100~Myr (from the bluest to the reddest 255$-$547 point); the
`knee' at 547$-$814$\simeq -$0.4 marks an age of about 10~Myr 
(Bruzual \& Charlot 1995).}
\end{document}